\RequirePackage{fix-cm}
\documentclass[twocolumn]{svjour3}          
\smartqed  
\usepackage{graphicx}
\usepackage{graphicx}
\usepackage{dcolumn}
\usepackage{bm}
\usepackage{hyperref}
\usepackage{graphicx}%
\usepackage{multirow}%
\usepackage{amsmath,amssymb,amsfonts}%
\usepackage{mathrsfs}%
\usepackage[title]{appendix}%
\usepackage{xcolor}%
\usepackage{textcomp}%
\usepackage{manyfoot}%
\usepackage{booktabs}%
\usepackage{algorithm}%
\usepackage{algorithmicx}%
\usepackage{algpseudocode}%
\usepackage{listings}%
\begin{document}

\title{Reexamination of the realtime protection for user privacy in practical quantum private query
}
\subtitle{}


\author{Chun-Yan Wei \and Xiao-Qiu Cai \and Tian-Yin Wang 
}

\institute{Chun-Yan Wei, Xiao-Qiu Cai, and Tian-Yin Wang \at
              School of Mathematical Science, Luoyang Normal University, Luoyang 471934, China\\
              \email{tianyinwang1979@163.com}    
}

\date{Received: date / Accepted: date}

\maketitle

\begin{abstract}
Quantum private query (QPQ) is the quantum version for symmetrically private retrieval. However, the user privacy in QPQ is generally guarded in the non-realtime and cheat sensitive way. That is, the dishonest database holder's cheating to elicit user privacy can only be discovered after the protocol is finished (when the user finds some errors in the retrieved database item). Such delayed detection may cause very unpleasant results for the user in real-life applications. Current efforts to protect user privacy in realtime in existing QPQ protocols mainly use two techniques, i.e., adding an honesty checking on the database or allowing the user to reorder the qubits. We reexamine these two kinds of QPQ protocols and find neither of them can work well. We give concrete cheating strategies for both participants and show that honesty checking of inner participant should be dealt more carefully in for example the choosing of checking qubits. We hope such discussion can supply new concerns when detection of dishonest participant is considered in quantum multi-party secure computations.
\end{abstract}

\section{Introduction}
\label{intro}
Symmetrically private information retrieval (SPIR) \cite{SPIR} is a fundamental secure two-party computation task in which a user Alice queries an item $x_i$ (generally assumed to be a bit) from a database $X=x_{1}x_{2}\cdots x_{N}$ without leaking the retrieval address $i$ (i.e., user privacy), while the database holder Bob hopes that Alice could obtain no more than the database item she queries (database security). Till now, many SPIR protocols have been designed in classical cryptography. However, with the progress in quantum algorithms, classical cryptography might be vulnerable to an adversary with quantum computers \cite{Shor,Grover}. Luckily, this defect can be overcome by quantum cryptography as its security is guarded by physical laws \cite{QC}.

Quantum private query is the quantum scheme for the SPIR task. However, the task of SPIR cannot be realized ideally even in quantum cryptography \cite{unideal}. More practically, QPQ slightly loosens the protection for the privacies of both sides. That is, Alice generally can elicit a few more bits than the ideal requirement (i.e., just 1 bit) from database, and the protection of user privacy is cheat-sensitive and non-realtime (that is, if Bob cheats to infer the retrieval address, he will supply false database item to Alice and hence can be detected with a nonzero probability after the retrieval is finished). Earlier QPQ protocols \cite{GLM,GLM1,O} utilizing unitary operations show great significance in theory, but they are not loss-tolerant and for large database the dimension of unitary operation will become too high to implement.

A more practical style of QPQ, i.e., QKD-based QPQ, was proposed in 2011 by Jakobi \emph{et al}. \cite{J}. Such QPQ generally contains a quantum stage and two classical steps. In the quantum stage, Alice and Bob shares a raw oblivious key $K_r$ which is known to Bob completely and to Alice partially, and Bob does not know which bits are known to Alice. In the classical steps, they first add the substrings of the raw key $K_r$ to obtain a final key $K_f$ so that Alice's knowledge on $K_f$ can be reduced to roughly one bit; then Bob uses $K^{f}$ to encrypt the database according to a shift announced by Alice so that she can extract the wanted item correctly from the encrypted database. QKD-based QPQ is more practical because it can tolerate the channel loss and can be implemented with current technology (many QKD protocols have become mature enough for the real-life applications). With this merit, QKD-based QPQ has attracted a great deal of concern and fruitful results have been achieved \cite{Gao2019}. Many QKD protocols have been exploited to realize QPQ, including interesting variants of SARG04 QKD \cite{J,SARG04,Gao2012,Yu2015} and BB84 QKD \cite{Wei2014}, counterfactual QKD \cite{Zhang2013}, two-way QKD schemes \cite{Wei2016}, round-robin differential-phase-shift QKD \cite{RRDPS,Liu2015} and so on. Meanwhile, remarkable progress has been made in the classical postprocessing \cite{Rao,Gao2015}, experimental implementation \cite{Chan,Wang2023,Kon2021}, strategies of resisting the joint-measurement attack \cite{Wei2016} and adapting the unideal source \cite{Wei2018,Liu2022}, channels \cite{Gao2015,Chan,Wei2020,Wang,Yang} as well as eliminating the security flaws of imperfect equipments in the device-independent or measurement-device-independent mode \cite{MDI1,DI,MDI2}.

Quite a few of above advancements in QKD-based QPQ aim to highlight the user privacy. The user privacy in QPQ is guarded in the non-realtime and cheat-sensitive way. That is, if Bob cheats, he may sends false database item to Alice, and Alice can find such cheating when she finds some errors in her obtained item on some moment after the finish of the retrieval. This delayed detection may cause very unpleasant results for the user in practical scenarios. Imagining that a famous stock broker may make a false purchase with the obtained incorrect stock item and the the leaking of his/her interest may cause serious losses. Current efforts to protect user privacy in realtime mainly use two techniques. One is adding an honesty checking on the database (HCD) to detect dishonest database holder in realtime, so this kind of QPQ can be called ``QPQ with HCD'' \cite{Yu2015,Zhou2018,Ye2020}; the other is allowing the user to reorder the qubits (ROQ) to prevent the database holder from eliciting user privacy, so this kind of QPQ can be called ``QPQ with ROQ''\cite{Zheng2019,Chang2019}.

As we know, preventing dishonest participants is significantly difficult than preventing outside eavesdroppers. On one hand, dishonest participants have the opportunities to transmit fake states, measure dishonestly, treat the checking and unchecking qubits differently or response more trickly, hence having great advantage in escaping from being detected than outside eavesdroppers. On the other hands, any strategy to improve the user privacy should be dealt more carefully, otherwise it may cause damage to database security because the two aspects are in a trade-off relationship. With these considerations, we reexamine the above two kinds of QPQs and find neither of them can really supply a realtime protection for user privacy. Worse yet, the database security will be damaged greatly. Concretely, in the QPQ with HCD, dishonest database can replace the honest measurement with two partial measurements, and then treats the checking qubits and the unchecking ones differently to obtain some advantage about user privacy; dishonest user can cheat to select the qubits contributing to  inconclusive raw key bits as the checking qubits so that after dropping them the proportion of conclusive bits can be increased, as a result, the user can obtain more database items than expected. In the QPQ with ROQ, dishonest Bob can elicit some advantage by checking the numbers of different measurement outputs published by Alice (though in an unknown rearranged order); dishonest Alice can store all qubits and measures them with correct basis after the honesty checking to obtain the whole raw key and the whole database. Therefore, neither of the two kinds of QPQs aiming to protect user privacy in realtime can work well. The honesty checking of participants in quantum secure computations should be dealt more carefully, otherwise it may harm the privacy of other parties.

The remainder of this paper is organized as follows. In Sects.2 and 3, we analyze the security of  ``QPQ with HCD'' and
 ``QPQ with ROQ'', respectively. Finally, we conclude in Sect.4.
\section{Cryptanalysis of QKD-based QPQ with HCD}
We first discuss the practical QPQ protocols which aim to detect an dishonest database holder in realtime with an honesty checking. To elicit user privacy, the database holder needs to know which raw key bits are known by the user, by sending fake states or conducting dishonest measurement and so on. To prevent the dishonest behaviours of database, the protocols of style ''QPQ with HCD'' generally ask the database to announce partial information about his/her measurement outputs or transmitted states and then check this announcement with some randomly chosen carrier qubits. Unfortunately, such honesty checking generally cannot work because the database holder can divide the honest operation defined by the protocol into partial ones to supply correct partial information in the announcement. We here take Yu et al.'s protocol as an example to show that the honest measurement of database can be split to two steps. As a result, dishonest database can supply correct announcement via one step before the honesty checking, then conducts the other step honestly for the checking qubits to supply correct answers and conducts other operation on the unchecking ones to elicit whether this raw key bit is known by the user.
\subsection{Review of Yu et al.'s protocol}

Yu et al.'s protocol contains the following 4 stages.

 \textbf{Stage 1. Generation of the oblivious key.} The database generates a key $b_1b_2 \cdots b_{kN}$, then steps (1a) (1b) (1c) are iterated over $i$ from $1$ to $kN$ so that the user can obtain partial key bits obliviously.
\begin{itemize}
\item \textbf{(1a)} The user sends the database one of $\{|0\rangle$, $|1\rangle$, $|+\rangle$, $|-\rangle\}$ randomly (in the original protocol the carrier qubits are written as $|\uparrow\rangle, |\downarrow\rangle, |\leftarrow\rangle, |\rightarrow\rangle$, and here we replace them with $|0\rangle, |1\rangle, |+\rangle, |-\rangle$ to maintain the consistency in expression).

\item \textbf{(1b)} The database measures the received state in Z basis $\{|0\rangle,|1\rangle\}$ (X basis $\{|+\rangle,|-\rangle\}$) when $b_i=0$ ($b_i=1$), and announces 0 (1) if the outcome state is $|0\rangle$ or $|+\rangle$ ($|1\rangle$ or $|-\rangle$).

\item \textbf{(1c)} The user derives $b_i$ with database's announcement and the original state he/she prepared. For example, when the prepared state is $|0\rangle$, if the database announces 1 in step (1b), the user knows that the database's measurement basis is $\{|+\rangle,|-\rangle\}$ and hence obtains an conclusive key bit $b_i=1$, otherwise the user obtains an inconclusive key bit.
\end{itemize}

\textbf{Stage 2. Honesty checking of database.} The user randomly selects a portion of locations where he/she has conclusive key bits and asks the database to announce the measurement outcomes of corresponding states. If he/she found the database cheating, the user aborts the protocol.

\textbf{Stage 3. Classical postprocessing.} After dropping the checking bits, the key $b_1b_2\cdots b_{kN}$ becomes $b_1b_2\cdots b_{k'N}$. Then they divide the key into substrings and then add them bitwise to create a final key $K$ so that the user knows  roughly one final key bit. If the user does not know any bit in the final key, the protocol restarts.

\textbf{Stage 4. Retrieval.} The user announces $s=j-i$ if he/she knows the $j$-th bit of $K$ and wants the $i$-th database item. The database holder shifts the key $K$ according to $s$, then adds it with the database, and finally sends the encrypted database to the user. The user elicits his/her wanted database item with the known bit in $K$.
\subsection{Cryptanalysis of Yu et al.'s protocol}
We here show the honesty checking cannot detect dishonest database holer, worse yet, it can supply dishonest user some chance to elicit more database items.
\subsubsection{Attack of dishonest database holder}
\begin{figure*}[htp]
\centering
\includegraphics[scale=0.5,trim=80 50 10 10, clip]{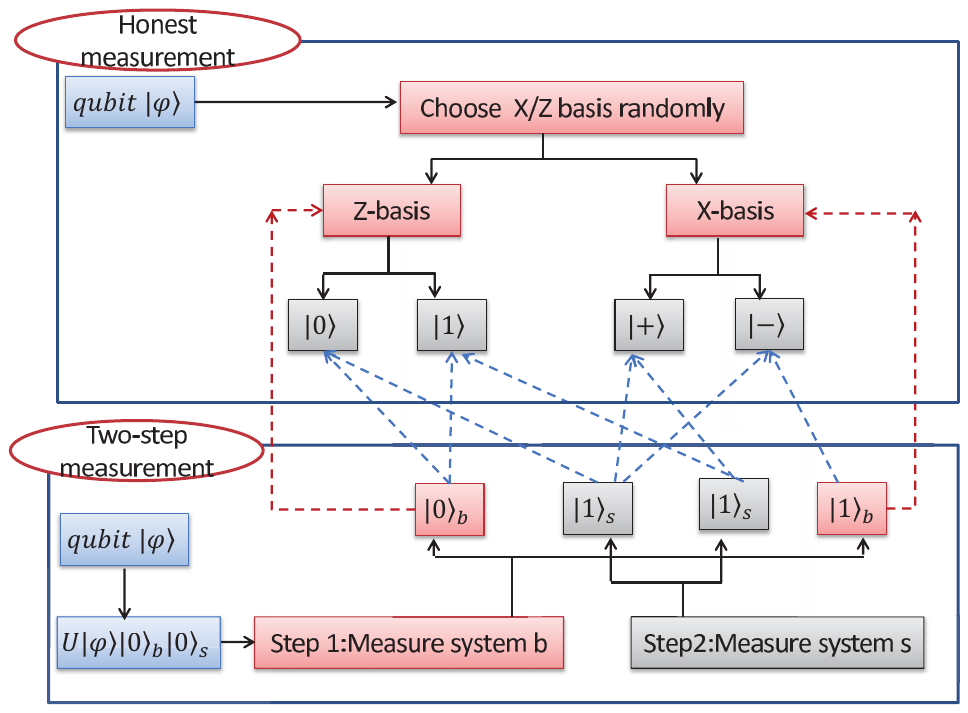}
\caption{\label{fig:epsart}The relationship between honest measurement and the two-step measurement.}
\end{figure*}

We now give a two-step measurement attack of dishonest database holder. By this attack, dishonest database holder not only can pass the realtime honesty checking in stage 2 but also can get advantage on eliciting user privacy. Concretely, for each received qubit $|\varphi\rangle_c \in \{|0\rangle, |1\rangle, |+\rangle |-\rangle\}$, the database prepares the state $|0\rangle_b|0\rangle_s$, and then conducts the unitary operation
\begin{equation}
\begin{split}
U & =|0\rangle_c\langle0|\otimes|0\rangle_b\langle0|\otimes I_s+|1\rangle_c\langle1|\otimes|0\rangle_b\langle0|\otimes  X_s\\&+|+\rangle_c\langle+|\otimes|1\rangle_b\langle1|\otimes I_s+|-\rangle_c\langle-|\otimes|1\rangle_b\langle1|\otimes X_s
\end{split}
\end{equation}
on $|\psi\rangle_c|+\rangle_b|0\rangle_s$, where $I$ is the identity operation and $X$ is the operator
$
\begin{pmatrix}
0 & 1\\
1 & 0
\end{pmatrix}.
$
Then we have
\begin{equation}
\begin{split}
&\quad U|0\rangle_c|+\rangle_b|0\rangle_s\\
&=\frac{1}{\sqrt{2}}|0\rangle_c|0\rangle_b|0\rangle_s+\frac{1}{2}|+\rangle_c|1\rangle_b|0\rangle_s+\frac{1}{2}|-\rangle_c|1\rangle_b|1\rangle_s\\
&=\frac{\sqrt{3}}{2}|\psi_1\rangle_{cb}|0\rangle_s+\frac{1}{2}|-1\rangle_{cb}|1\rangle_s,
\end{split}
\end{equation}
\begin{equation}
\begin{split}
&\quad U|1\rangle_c|+\rangle_b|0\rangle_s\\ &=\frac{1}{\sqrt{2}}|1\rangle_c|0\rangle_b|1\rangle_s+\frac{1}{2}|+\rangle_c|1\rangle_b|0\rangle_s-\frac{1}{2}|-\rangle_c|1\rangle_b|1\rangle_s\\
&=\frac{\sqrt{3}}{2}|\psi_2\rangle_{cb}|1\rangle_s+\frac{1}{2}|+1\rangle_{cb}|0\rangle_s,
\end{split}
\end{equation}
\begin{equation}
\begin{split}
&\quad U|+\rangle_c|+\rangle_b|0\rangle_s\\ &=\frac{1}{2}|0\rangle_c|0\rangle_b|0\rangle_s+\frac{1}{2}|1\rangle_c|0\rangle_b|1\rangle_s+\frac{1}{\sqrt{2}}|+\rangle_c|1\rangle_b|0\rangle_s\\
&=\frac{\sqrt{3}}{2}|\psi_3\rangle_{cb}|0\rangle_s+\frac{1}{2}|10\rangle_{cb}|1\rangle_s,
\end{split}
\end{equation}
\begin{equation}
\begin{split}
&\quad
U|-\rangle_c|+\rangle_b|0\rangle_s\\ &=\frac{1}{2}|0\rangle_c|0\rangle_b|0\rangle_s-\frac{1}{2}|1\rangle_c|0\rangle_b|1\rangle_s+\frac{1}{\sqrt{2}}|-\rangle_c|1\rangle_b|1\rangle_s\\
&=\frac{\sqrt{3}}{2}|\psi_4\rangle_{cb}|1\rangle_s+\frac{1}{2}|00\rangle_{cb}|0\rangle_s,
\end{split}
\end{equation}
where
\begin{equation*}
|\psi_1\rangle=\frac{\sqrt{2}}{\sqrt{3}}|00\rangle_{cb}+\frac{1}{\sqrt{3}}|+1\rangle_{cb},
\end{equation*}
\begin{equation*}
|\psi_2\rangle=\frac{\sqrt{2}}{\sqrt{3}}|10\rangle_{cb}-\frac{1}{\sqrt{3}}|-1\rangle_{cb},
\end{equation*}

\begin{equation*}
|\psi_3\rangle=\frac{\sqrt{2}}{\sqrt{3}}|+1\rangle_{cb}+\frac{1}{\sqrt{3}}|00\rangle_{cb},
\end{equation*}

\begin{equation*}
|\psi_4\rangle=\frac{\sqrt{2}}{\sqrt{3}}|-1\rangle_{cb}-\frac{1}{\sqrt{3}}|10\rangle_{cb}.
\end{equation*}
In this case, the honest measurement of database, i.e., measuring the qubit randomly in $X$ or $Z$ basis, can be simulated by a two-step measurement (see Fig.1). That is, measuring system $b$ is equivalent to choosing the measurement basis randomly (the random output $|0\rangle_b$ ($|1\rangle$) corresponds to choosing the Z (X) basis), and measuring system $s$ is equivalent to measure the qubits with the basis defined by system $b$. More importantly, the two steps can be conducted in arbitrary order.
Then in stage (1b) the database holder can measure system $s$ in basis $\{|0\rangle,|1\rangle\}$ and announces a bit 0(1) if the output is $|0\rangle$ ($|1\rangle$). Then in stage 2, if this qubit is selected to check the honesty of database, the database can measure the system $b$ with basis $\{|0\rangle,|1\rangle\}$ and replies with $|0\rangle$ ($|1\rangle$, $|+\rangle$ or $|-\rangle$) when the output of systems b and s is $|0\rangle_b|0\rangle_s$ ($|0\rangle_b|1\rangle_s$, $|1\rangle_b|0\rangle_s$ or $|1\rangle_b|1\rangle_s$). Clearly, the database can pass the honesty checking successfully.

Now we estimate the advantage the database can achieve by this attack. After dropping the checking qubits in stage 2, the dishonest database still holds systems $c$ and $b$ for the remaining qubits, and he/she can use them to elicit whether corresponding raw key bit is known by the user. Without loss of generality, we consider the case that the database's output is $|0\rangle$ and hence he/she announces 0 in stage(1b). He/she can infer from eqs.(2-5) that, the user obtains a conclusive bit when the original state of received qubit is $|1\rangle$ ($|-\rangle$) and corresponding systems $c$ and $b$ collapse to $|+1\rangle_{cb}$ ($|00\rangle_{cb}$); and the user obtains an inconclusive bit when the original state of received qubit is $|0\rangle$ ($|+\rangle$) and corresponding systems $c$ and $b$ collapse to the state $|\psi_1\rangle$ ($|\psi_3\rangle$). That is, when the database's measurement output of system $s$ is $|0\rangle$, if the user obtains a conclusive raw key bit (which happens with probability $\frac{1}{4}$), systems $c$ and $b$ will collapse to the state
\begin{equation}
\rho_{c}=\frac{1}{2}(|+1\rangle\langle+1|+|00\rangle\langle00|);
\end{equation}
if the user obtains an inconclusive one (which happens with probability $\frac{3}{4}$), systems $c$ and $b$ will collapse to the state
\begin{equation}
\rho_{in}=\frac{1}{2}(|\psi_1\rangle\langle\psi_1|+|\psi_3\rangle\langle\psi_3|).
\end{equation}
In this case, the database infer whether the user obtains a conclusive bit by discriminating the two mixed states $\{\frac{1}{4},\rho_c;\frac{3}{4},\rho_{in}\}$. As the two mixed states have the same supports, they cannot be discriminated unambiguously, while the minimum error probability to discriminate them is 0.1464 \cite{qufen1,qufen2}. That is, dishonest Bob can elicit the user's conclusiveness of each raw key with error probability 0.1464, equal with that of Jakobi et al.'s protocol\cite{J}.

Therefore, the realtime honesty checking of database holder in Yu et al.'s protocol cannot improve the user privacy because it cannot detect the above cheating of database and cannot reduce dishonest database's advantage compared with Jakobi et al.'s protocol \cite{J}. The main reason is that, database can announce correctly before honesty checking by partial measurement instead of by the expected honesty measurement, which gives him/her the chance to treat the checking qubits and unchecking ones differently to pass the checking and elicit user privacy simultaneously.

\subsubsection{Attack of dishonest user}

We now turn to another serious flaw of the involved realtime checking, that is, it may greatly hurt the database security. Note that, to detect dishonest database, the user needs to choose the qubits contributing to his/her conclusive key bits as the checking qubits. However, if the user cares not his/her privacy but the amount of obtained database, he/she may give up the chance of detecting dishonest database and tries to elicit more database items via the honesty checking. Concretely, the user can select those contributing to his/her inconclusive bits as the checking qubits, corresponding inconclusive key bits will be dropped from the raw key and the proportion of his/her conclusive bits will increase significantly (note that, the checking qubits generally occupy a large proportion of the transmitted qubits, e.g., 50\% as required in ref.\cite{Zhou2018}). As a result, the user can gain many more database items than expected (even the whole database if the proportion of checking qubits achieves 3/4).

Therefore, the involved realtime honesty checking of database cannot supply realtime protection for user privacy. Worse yet, it may cause great damage to database security. Therefore, the honesty checking of inner participants in quantum secure computations should be handled more carefully, in for example the choosing of checking qubits.

\section{Cryptanalysis of QKD-based QPQ with ROQ }
We now turn to another kinds of QPQ protocols which aim to protect user privacy in realtime via reordering of qubits. We show these protocols cannot work well by examining Chang et al.'s protocol.

\subsection{Review of Chang et al.'s protocol with ROQ}\label{sec2}
Chang et al.'sl uses two-way communication, that is, the database holder Bob sends the qubits to the user Alice, and Alice measures the received qubits and sends them back in a new order (unknown to Bob) to prevent Bob from eliciting which raw key bits are known by Alice. The concrete protocol is as follows.

\textbf{Step 1.} The database holder Bob sends the user Alice a sequence of qubits which are randomly in one state of $\{|0\rangle,|1\rangle, |+\rangle, |-\rangle\}$.

\textbf{Step 2.} For each received qubit, Alice measures it in Z basis with probability $\eta$ or in X basis with probability $1-\eta$. Then she makes every $n$ ($n\geq 4$) particles into a group, rearranges the order of them within each group, and sends all the rearranged particles back to Bob. Finally she tells Bob the measurement bases and outputs of the reordered qubits within each group and keeps the new orders secret.

\textbf{Step 3.} Bob checks whether Alice increases $\eta$ or not. Concretely, Bob measures each received particle according to the basis Alice tells him and compares his outputs with those published by Alice. Once he finds that Alice increases the value of $\eta$, he aborts the protocol.

\textbf{Step 4.} Bob checks whether Alice sends him fake qubits or not. Concretely, Bob publishes the locations of all X-basis particles (i.e., $|+\rangle, |-\rangle$) in the original sequence prepared in step 1 and requires Alice to disclose the rearranged positions of these particles. By checking these qubits, Bob can determine whether Alice sends him fake qubits or not. If the qubits sent back to Bob are judged as fake qubits, the protocol aborts.

\textbf{Step 5.} They discard the checking qubits in step 4, then all the remaining ones are prepared in Z-basis originally. Following their order in the original sequence, Alice (Bob) records the measurement outputs (the original states) of them. Concretely, $|0\rangle$ $(|1\rangle)$ is recorded as bit 0 (1), and if Alice measured the qubit in X-basis in step 2, she records an inconclusive bit ``?'' here. In this way, they share a raw key $K_{Raw}$ which is known to Bob totally and to Alice partially.

\textbf{Step 6.} Alice and Bob post-process the raw key by bitwise adding the substrings of $K_{Raw}$ to obtain a final key $K$ so that Alice knows roughly one bit in  $K$.

\textbf{Step 7.} If Alice knows the $j$-th bit in $K$ and wants the $i$-th item from the database, she announces the value $s=i-j$. Bob encrypts the database with the final key $K$, shifted by $s$, and then sends the encrypted database to Alice. Finally, Alice acquires her wanted database item with her known bit in $K$.

\subsection{Cryptanalysis of Chang et al.'s protocol}\label{sec3}
\subsubsection{Attack of dishonest database holder Bob}
We now show that, Bob can obtain some advantage from the measurement outputs published by Alice even though he has no knowledge on Alice's rearrangement of qubits, which clearly harms user privacy. Note that in Chang et al.'s protocol Bob knows the numbers of $|0\rangle$, $|1\rangle$, $|+\rangle$ and $|-\rangle$ in Alice's measurement outputs as well as in the original states within each $n$-qubit group. It will supply him some advantage to elicit user privacy. We take $n=6$ (note that Chang et al.'s protocol sets $n\geq 4$) as an example. Suppose that in one group the six original qubits sent by Bob is $|0\rangle$, $|1\rangle$, $|1\rangle$, $|1\rangle$, $|+\rangle$, $|-\rangle$ and the measurement outputs published by Alice (after the rearrangement) are $|1\rangle$, $|+\rangle$, $|+\rangle$,$|1\rangle$, $|-\rangle$, and $|-\rangle$ (i.e. without $|0\rangle$), then Bob knows that the qubit $|0\rangle$ he sent is measured in X basis by Alice and hence it contributes an inconclusive raw key bit for her. If this raw key bit contributes to the $i$-th final key bit after the bitwise adding, Bob knows that this final key bit will not be used to encrypt the database item Alice wants, therefore he can reduce the scope of retrieval address and gain virtual benefit in obtaining user privacy.

\subsubsection{Attack of dishonest user Alice}
We now show that dishonest Alice in Chang et al.'s protocol can obtain the whole database, which means that the database security can be broken completely. The concrete strategy is as follows.
\begin{itemize}

\item{In step 2, Alice divides the received qubits into $n$-qubit groups, and stores them in her register. Then for each group, instead of measuring the qubits and sending them back to Bob in a new order, she sends Bob a fake $n$-qubit sequence in which the proportions of $|0\rangle$, $|1\rangle$, $|+\rangle$ and $|-\rangle$ are $\frac{\eta}{2}$, $\frac{\eta}{2}$, $\frac{1-\eta}{2}$, and $\frac{1-\eta}{2}$, respectively (obviously, this faked qubit sequence can pass Bob's checking about $\eta$ in step 3). Then she tells Bob the states of this sequence of qubits instead of the measurement outputs of the reordered carrier qubits she receives.}

\item{Then in step 4, when Bob announces the positions of the X-basis particles, Alice measures corresponding particles stored in her register and replies as follows. If the measurement output is $|+\rangle$, she checks her fake sequence and selects a qubit in state $|+\rangle$ with probability $1-\eta$ or a state in $|0\rangle$ or $|1\rangle$ with probability $\eta$, and announces the location of this qubit. Similarly, if the measurement output is $|-\rangle$, she checks her fake sequence and selects a qubit in state $|-\rangle$ with probability $1-\eta$ or qubit in state $|0\rangle$ or $|1\rangle$ with proability  $\eta$, and announces the location of this qubit. Clearly, this strategy can pass Bob's checking in step 4.}

\item{In step 5, after dropping the X-basis particles, Alice measures the remaining qubits stored in her register in Z basis and hence obtains the whole raw key $K_{Raw}$.}

\item{As a result, she can obtain the whole final key $K$ in step 6 and the whole database in step 7.}
\end{itemize}

The main reason for the success of above attack is that the locations of the reordered qubits are not bounded, that is, even for the fake qubit sequence sent to Bob in step $2$, Alice can change the locations of checking qubits according to her measurement outputs to pass the honesty checking in step 4. As a result, dishonest Alice can store all carrier qubits and postpone to measure them in correct basis after the honesty checking and obtains the whole raw key as well as the whole database.

In conclusion, though the rearrangement of qubits supplies some protection for user privacy, it also supplies dishonest user Alice a chance to destroy database security completely, which is unacceptable for database holder.

\section{Conclusions}

We reexamine the existing practical QPQ protocols aiming to supply realtime protection for user privacy and show that preventing dishonest participant from cheating is a troublesome task. Concretely, inner participant generally has great potential to escape being detected via sending fake state, measuring dishonesty, announcing more tricky and so on. On the other hand, the participant dominating the honesty checking may obtain virtual advantage by for example choosing special checking qubits. We hope such discussion can supply some new concerns for detecting the dishonest participant in quantum multi-party secure computations.

\begin{acknowledgements}
This work was supported by the Natural Science Foundation of Henan Province (Grant No. 212300410062), the National Natural Science Foundation of China (Grant Nos. 61902166, 62172196, 62272208).
\end{acknowledgements}



\end{document}